\documentclass[twocolumn]{aastex6}
\begin{document}

\title{An absence of radio-loud active galactic nuclei in geometrically flat quiescent galaxies: implications for maintenance-mode feedback models}
\author{
Ivana Bari\v{s}i\'c$^1$,
Arjen van der Wel$^{1,2}$,
Josha van Houdt$^{1}$,
Michael V. Maseda$^{3}$,
Eric F. Bell$^{4}$,
Rachel Bezanson$^{5}$,
Yu-Yen Chang$^{6}$,
Huub R\"{o}ttgering$^{7}$,
Glenn van de Ven$^{8}$,
Po-Feng Wu$^{1}$}

\thanks{$^1$Max-Planck Institut f\"ur Astronomie, K\"onigstuhl 17, D-69117, Heidelberg, Germany}
\thanks{$^2$Sterrenkundig Observatorium, Department of Physics and Astronomy, Ghent University, Belgium}
\thanks{$^3$Leiden Observatory, Leiden University, P.O.Box 9513, NL-2300 AA Leiden, The Netherlands}
\thanks{$^5$University of Pittsburgh, Department of Physics and Astronomy, 100 Allen Hall, 3941 OHara St, Pittsburgh PA 15260, USA}
\thanks{$^4$Department of Astronomy, University of Michigan, 1085 S. University Ave, Ann Arbor, MI 48109, USA}
\thanks{$^{6}$Academia Sinica Institute of Astronomy and Astrophysics, No.1 Section 4 Roosevelt Rd., 11F of Astro-Math Building, Taipei 10617, Taiwan}
\thanks{$^7$Leiden Observatory, Leiden University, P.O. Box 9513, NL-2300 RA Leiden, The Netherlands}
\thanks{$^{8}$European Southern Observatory, Karl-Schwarzschild-Straße 2, 85748 Garching bei Munchen, Germany}
%\thanks{$^2$University of Pittsburgh, Department of Physics and Astronomy, 100 Allen Hall, 3941 OHara St, Pittsburgh PA 15260, USA}
%\thanks{$^3$Astrophysics Science Division, Goddard Space Flight Center, Code 665, Greenbelt, MD 20771, USA}
%\affil{TBD}
%\thanks{$^5$European Southern Observatory, Alonso de C\'{o}rdova 3107, Casilla 19001, Vitacura, Santiago, Chile}
%\thanks{$^6$Leiden Observatory, Leiden University, P.O.Box 9513, NL-2300 AA Leiden, The Netherlands}
%\thanks{$^7$Department of Physics, Faculty of Science, University of Zagreb,  Bijeni\v{c}ka cesta 32, 10000  Zagreb, Croatia}
%\thanks{$^{10}$Instituto de Astrof\'{i}́sica e Ci\'{e}ncias do Espa\c{c}̧o, Universidade de Lisboa, OAL, Tapada da Ajuda, P-1349-018 Lisboa, Portugal}
%\thanks{$^{8}$Department of Astronomy, University of Michigan, 1085 S. University Ave, Ann Arbor, MI 48109, USA}
%\thanks{$^9$Space Telescope Science Institute, 3700 San Martin Drive, Baltimore, MD 21218, USA}
%\thanks{$^10$Department of Physics, Lancaster University, Lancaster LA1 4YB, UK}
%\thanks{$^{11}$Department of Astronomy, Yale University, New Haven, CT06511, USA}
%\thanks{$^{12}$INAF-Osservatorio Astrofsico di Arcetri, Largo Enrico Fermi 5, I-50125 Firenze, Italy}
%\thanks{$^{13}$Department of Physics and Astronomy, York University, 4700 Keele Street, Toronto, Ontario, ON MJ3 1P3, Canada}\\

\email{barisic@mpia.de}

%\maketitle

\begin{abstract}

Maintenance-mode feedback from low-accretion rate AGN, manifesting itself observationally through radio-loudness, is invoked in all cosmological galaxy formation models as a mechanism that prevents excessive star-formation in massive galaxies (M$_*$ $\gtrsim$ 3$\times$10$^{10}$ M$_{\odot}$). We demonstrate that at a fixed mass the incidence of radio-loud AGN (L $>$ 10$^{23}$ WHz$^{- 1}$) identified in the FIRST and NVSS radio surveys among a large sample of quiescent (non-star forming) galaxies selected from the SDSS is much higher in geometrically round galaxies than in geometrically flat, disk-like galaxies. As found previously, the RL AGN fraction increases steeply with stellar velocity dispersion $\sigma_*$ and stellar mass, but even at a fixed velocity dispersion of 200-250 kms$^{-1}$ this fraction increases from 0.3\% for flat galaxies (projected axis ratio of q $<$ 0.4) to 5\% for round galaxies (q $>$ 0.8). We rule out that this strong trend is due to projection effects in the measured velocity dispersion. The large fraction of radio-loud AGN in massive, round galaxies is consistent with the hypothesis that such AGN deposit energy into their hot gaseous halos, preventing cooling and star-formation. However, the absence of such AGN in disk-like quiescent galaxies -- most of which are not satellites in massive clusters, raises important questions: is maintenance-mode feedback a generally valid explanation for quiescence; and, if so, how does that feedback avoid manifesting at least occasionally as a radio-loud galaxy? 

%Radio-loud active galactic nuclei (AGN) are invoked in all cosmological galaxy formation models to provide feedback that prevents further star formation in massive galaxies (M$_*$ $\gsim$ 3$\times$10$^{10}$ ). We demonstrate that the incidence of radio-loud AGN (L $>$ 10$^{23}$ W Hz$^{-1}$) among quiescent (non-star forming) galaxies is much higher in geometrically round galaxies than in flat (disk-like) galaxies, even after matching for stellar velocity dispersion, stellar mass and super-massive black hole mass. For round galaxies the fraction of radio-loud AGN rapidly rises from $\sim$0.1\% for galaxies with stellar velocity dispersion values $\sigma_*$ $<$100 km s$^{-1}$ to $>$10\% for galaxies with $\sigma_*$ $>$250 km s$^{-1}$. For disk-like quiescent galaxies this fraction is $<$0.1\% up to $\sigma_*$ $\sim$250 km s$^{-1}$ and we find upper limits of 0.2$-$0.5\% for $\sigma_*$ $>$250 km s$^{-1}$. %In fact, our analysis implies that among the $\sim$63,500 disk-like quiescent galaxies with $\sigma_*$ $>$150 km s$^{-1}$ in the Sloan Digital Sky Survey not a single one hosts a radio-loud AGN {\color{blue}{\textbf{(I don't see this mentioned anywhere in the text)}}}. 
%We conclude that the empirical evidence for AGN-driven maintenance-mode feedback is much stronger for massive, intrinsically round galaxies than for typical, disk-like quiescent galaxies.
\end{abstract}

\section{Introduction}

To explain the mismatch between the theoretical dark matter halo mass function and the observed galaxy stellar mass function, semi-analytical and hydrodynamical simulations of galaxy formation and evolution have implemented a set of feedback mechanisms to reduce the efficiency of star-formation; they rely on supernova feedback in low halo mass systems, and  super-massive black-hole (SMBH) feedback in systems with high mass halos \citep{croton06, bower06, fabian12, vogelsberger14, pillepich17}.
The most common picture states that the SMBH feedback generally manifests through radio-loudness in active galactic nuclei (AGN) in the form of the jets (``radio-mode'') or gas outflows (``quasar-mode'') as a consequence of a low- or high-accretion rate (respectively) of the gas onto the SMBH. 
This feedback picture states that powerful jets transfer their energy to the surrounding gas, heating it up and therefore preventing it from cooling and forming new stars.
``Radio-mode'' feedback has thus been introduced as the leading mechanism in explaining the mass growth deficiency in the most massive early type galaxies with SMBH. % -- the radio-loud fraction is therefore expected to be higher among quiescent then among star-forming galaxies at a fixed BH mass. 

Direct observational evidence for this picture in local galaxies comes from observations of cavities in the X-ray emitting gas caused by jets,  connecting the presence of radio-loud (RL) AGN with the absence of star-formation \citep{mcnamara07, heckman14}.
Indirect evidence in the local universe has also provided additional support to this radio-mode feedback picture \citep{matthews64, kauffmann03}, including studies that show a strong, increasing trend in the RL fraction with M$_*$ and stellar velocity dispersion $\sigma_*$ \citep{best05}.
According to the \cite{vdbosch16} BH mass $-$ $\sigma_*$ relation \citep[see also][]{gebhardt03, beifiori12}, this implies an increasing trend of the RL fraction with the BH mass, suggesting the importance of the BH mass as a parameter in setting the probability for a galaxy to become RL \cite[see also][]{haring04, terrazas17}.

In this work we explore how the incidence rate of RL AGN varies with geometric shape (flat or disk-like vs. round). 
We test this by examining the correlation of the RL fraction with $\sigma_*$ and the projected axis ratio for a sample of galaxies drawn from the Sloan Digital Sky Survey \citep[SDSS;][]{york00}, NRAO VLA Sky Survey \citep[NVSS;][]{condon98} and Faint Images of the Radio Sky at Twenty centimeter \citep[FIRST;][]{becker95} survey, and we present evidence that round galaxies host RL AGN much more frequently than disk-like galaxies at any fixed $\sigma_*$ and M$_*$. % and BH mass.
According to the 'radio-mode' feedback picture, and the association of the presence of RL AGN with quiescence, this raises a question to why are these galaxies not forming new stars in the absence of a visible heating source.

In Section 2 we give an overview of the data-set we use in this study, and in Section 3 we present or results and discuss the implications. In section 4 we give a brief summary of our main results.
We use standard cosmology: $H_0$ = 70(km/s)/Mpc, $\Omega_M$ = 0.3, $\Omega_{\Lambda}$ = 0.7 .

\section{Data and sample selection}
\label{data}
%In this section we introduce the data-sets analyzed in this work. We describe the sample selection criteria, and the method used to classify the sample among quiescent and star-forming galaxies. We also describe the method and criteria applied in selecting the radio-loud sub-sample. 

\subsection{SDSS}
The SDSS survey \citep{york00} provides unique and high quality information on various physical properties of a large number of present-day galaxies. Physical properties of those galaxies are measured and derived using photometry and spectroscopy, and include $\sigma_*$\footnote{http://classic.sdss.org/dr7/algorithms/veldisp.html}, star-formation rate (SFR), M$_*$, redshift and the projected axis ratio. The resolution of the spectroscopic data-set is $R$ = 1850 $-$ 2200, covering the wavelength range between 3800 \AA \ to 9200 \AA , with a minimum signal-to-noise S/N $\sim$ 14 \AA$^{-1}$. 

In this study we make use of the SFRs and M$_*$ obtained from modelling galaxy spectral energy distributions \cite[SED;][]{chang15} based on SDSS Data Release 7 \citep[DR7;][]{abazajian09} and WISE \citep{cutri13}. % The stellar masses are derived by comparing the information extracted from the spectra with that of the broad band photometry. 

%In order to create a mass-complete sample data set for this analysis, we apply a redshift based mass-cut (log(M$_*$) $>$ 10.6 + 2.28$\times$log($z$/0.1)) following \cite{chang15}. Additionally, we require a flux density measurement in the {\bf{rest-frame}} $u$-, $r$-, and $z$-band and a measured projected axis ratio. Combining these criteria, the number of galaxies from the original sample (601,169; SDSS DR7) {\bf{reduces}} down to 303,785 galaxies which {\bf{consequently}} span a redshift range from 0.02 $<$ z $<$ 0.14. {\bf{We note that the  $u$-, $r$-, and $z$-band measurement requirement reduces the sample size by less then 0.1\%, and is necessary in order to apply a color classification.}}
In order to create a volume-selected sample (303,785\footnote{ We removed 11058 galaxies (3.5\%) that lack rest-frame u-, r-, and z-band photometry and a measurement of the axis ratio.}) complete in M$_*$ we apply a redshift-dependent mass-cut: log(M$_*$) $>$ 10.6 + 2.28$\times$log($z$/0.1)) \citep{chang15}, consequently spanning a redshift range from 0.02 $<$ z $<$ 0.14.
For sample classification between star-forming and quiescent galaxies, we apply a color-based criteria following \cite{chang15}  ($u$ - $r$ $>$ 1.6 $\times$ ($r$ - $z$) + 1.1) for quiescent galaxies. Star-forming galaxies are considered to be all other galaxies that fall outside of this color-based criteria.

\subsection{NVSS and FIRST}

The NVSS \citep{condon98} 1.4 GHz sky survey north of $\delta= -40^{o}$ has been conducted %using the Very Large Array (VLA) radio interferometer 
in the period between 1993 and 1996. The angular resolution of the observed data-set is 45$\arcsec$, with the noise level rms of 0.45 mJy beam$^{-1}$ and the point source flux density 50\% completeness level at 2.5 mJy. For more details see \cite{condon98}.
The FIRST \citep{becker95} 1.4 GHz survey of the northern Galactic cap has been conducted in the period between 1993 and 2004. The average rms of the images is 0.15 mJy, with an angular resolution of 5$\arcsec$ and the flux density limit of 1 mJy.
Both surveys have been conducted using the Very Large Array (VLA) radio interferometer.

Taking advantage of the high resolution FIRST data-set, \cite{best12} cross-match sources with galaxies in the SDSS DR7. Flux density measurements are subsequently performed on the NVSS data-set. The resulting cross-matched catalog, with a limiting NVSS flux density value of 5 mJy, is complete for radio-loud AGN down to L$_{1.4GHz}$ $\sim$ 10$^{23}$ WHz$^{-1}$ at $z=0.1$. Since the sample is incomplete below L$_{1.4GHz}$ $<$ 2.5 $\times$ 10$^{23}$ WHz$^{-1}$ at highest redshifts, we apply a volume-correction that reaches at most a factor 4 for L$_{1.4GHz}$ $\sim$ 10$^{23}$ WHz$^{-1}$ at z = 0.14. 
%The NVSS \citep{condon98} 1.4 GHz sky survey north of $\delta$= -40$^{o}$ has been conducted using the Very Large Array (VLA) radio interferometer in the period between 1993 and 1996. The angular resolution of the observed data set is 45$\arcsec$, with the total intensity rms of 0.45 mJy beam$^{-1}$ and the point source flux density limited down to 2.5 mJy. For more details see \cite{condon98}.

\subsection{Selection of Radio-Loud AGN Hosts}
\begin{figure}
    \centering
    \includegraphics[width=\linewidth]{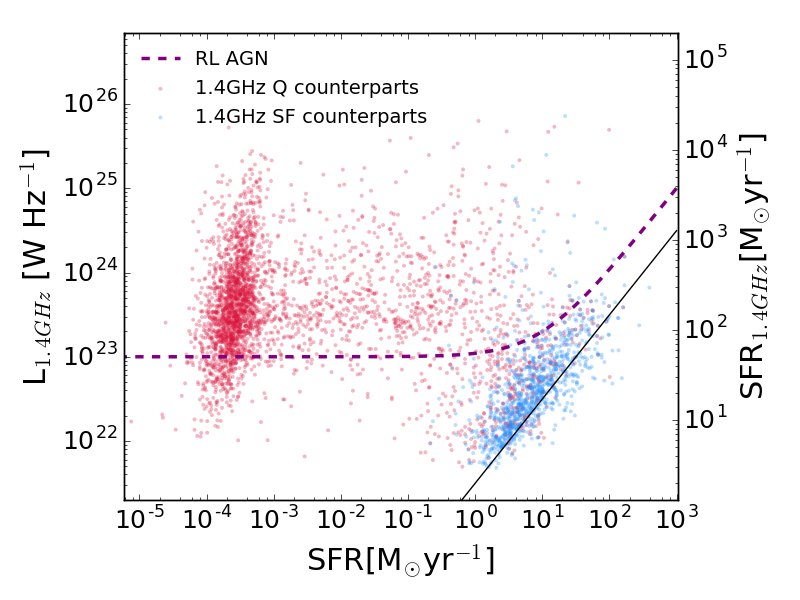}
    \caption{1.4 GHz luminosity vs. mid-IR SFR estimates \citep{chang15} for quiescent (red) and star-forming (blue symbols) galaxies. Right-hand side y-axis converts the 1.4 GHz luminosity into SFR. All data points have a radio counterpart in the NVSS, and the purple line indicates selected radio-loud (RL) AGN. The selection reflects the minimum required AGN luminosity of 10$^{23}$ WHz$^{-1}$ and accounts for a contribution of star-formation (see text for details).  The clump of galaxies at low SFR reflects the minimum SFR allowed in the SED model used by \citet{chang13} and has no physical meaning.}
    \label{fig:sample}
\end{figure}

To select a radio-loud sub-sample we consider galaxies with detected radio emission at 1.4 GHz by the NVSS and FIRST. From the total number of galaxies in the selected galaxy sample, the number of those detected with F$_{1.4 GHz}$ $>$ 5 mJy is 4,571, and they are shown with red and blue symbols in Figure~\ref{fig:sample}.

We use \cite{condon92} luminosity relation ($\alpha$ = - 0.8), to convert the NVSS measured flux density values F$_{1.4GHz}$ into luminosities. Derived luminosities at 1.4 GHz are then converted to radio-based SFRs using the \cite{bell03} radio$-$far-infrared calibration. 

We follow \cite{best05} 1.4 GHz luminosity limit L$_{1.4 GHz}$ $>$ 10$^{23}$ WHz$^{-1}$ to select radio-loud galaxies. We subtract the contribution of star-formation as follows: L$_{1.4GHz}$ - (3 $\times$ L$_{1.4GHz, SFR}$) $>$ 10$^{23}$ , where L$_{1.4GHz}$ is the luminosity based on the 1.4 GHz flux density value, and L$_{1.4GHz, SFR}$ is the expected L$_{1.4GHz}$ luminosity based on the SFR (taken from \cite{chang15}). In order to select RL AGN with high confidence we multiply L$_{1.4GHz, SF}$ by a factor 3 to account for uncertainties in radio- and mid-IR based SFR estimates.

%In addition to the \cite{best05} radio-loud galaxy selection criteria, we allow for a factor of 3 scatter in 1.4 GHz-based star-formation rates compared to the SED-based star-formation rates.
We keep the color-based classification from \cite{chang15} to separate the radio-detected sample between quiescent (red) and star-forming (blue symbols) galaxies. The selected radio-loud AGN sub-sample is indicated with the purple dashed line in Figure~\ref{fig:sample}.

%is shown with red symbols in Figure~\ref{fig:sample}, and a red dashed curve represents the combined radio-loud sub-sample selection criteria.

\begin{figure*}
    %\centering
    \includegraphics[width=\linewidth]{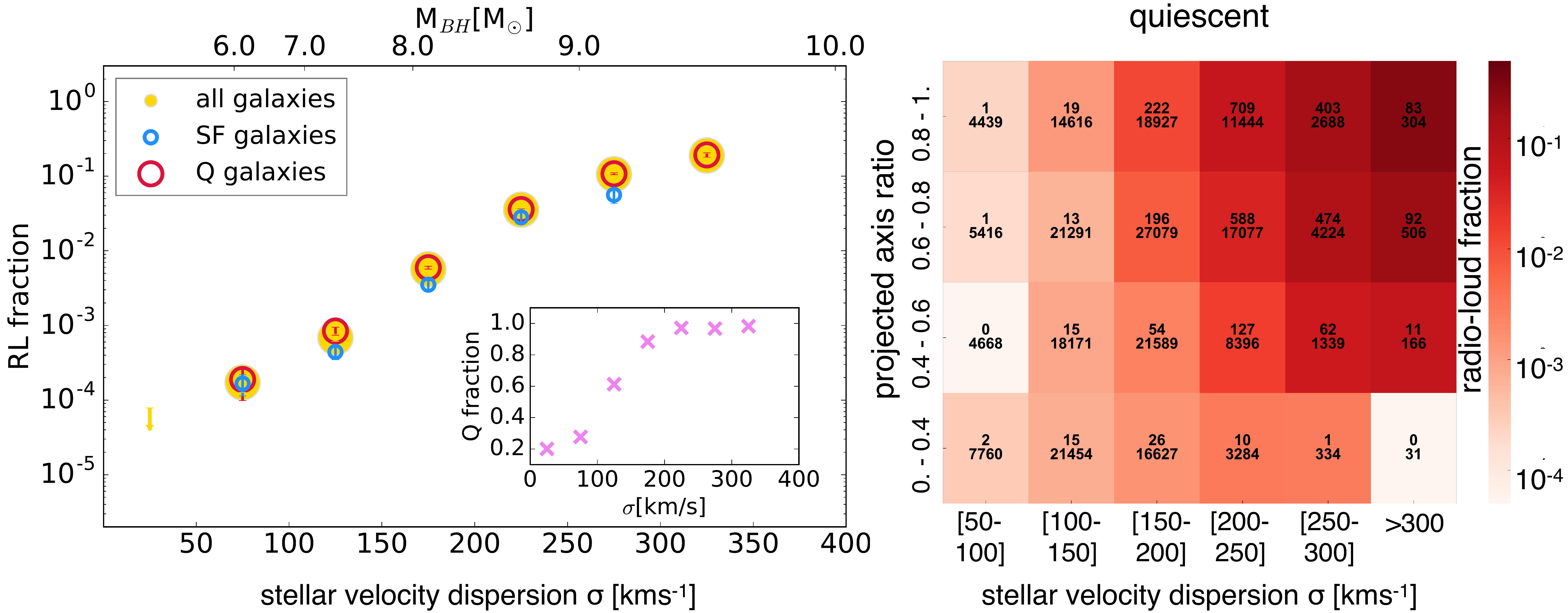}
    \caption{Left: RL fraction among all (yellow), quiescent (red empty) and star-forming (blue empty circles) galaxies as a function of the $\sigma_*$,  (including Poissonian error bars) and derived BH masses \cite{vdbosch16} on top. Yellow arrow centered at 25 kms$^{-1}$ represents RL fraction upper limit. The RL fraction increases with the $\sigma_*$ for both quiescent and star-forming galaxies. Violet crosses represent the fraction of quiescent galaxies among the total sample. Right: The heat-map of the observed axis ratio distribution as a function of the $\sigma_*$ for quiescent galaxies, color-coded by the RL fraction. We see an increasing trend in the RL fraction with $\sigma_*$ and with the projected axis ratio.}
    \label{fig:rlfrac}
\end{figure*}

\begin{figure*}
    \centering
    \includegraphics[width=\linewidth]{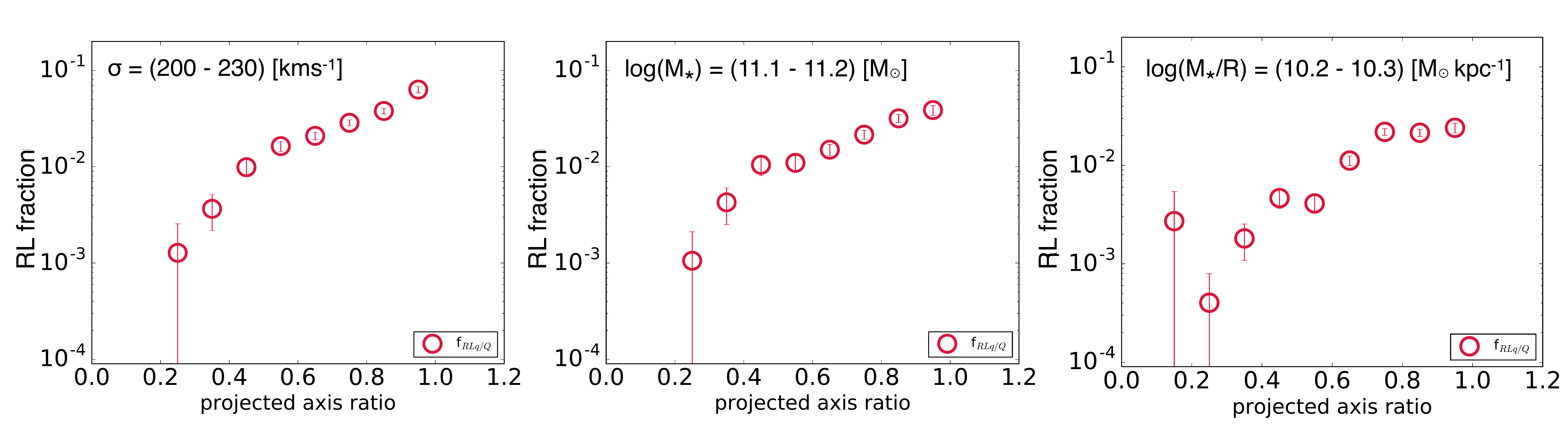}
    \caption{RL fraction among quiescent galaxies as a function of the projected axis ratio in the narrow bins of the $\sigma_*$ (left),  M$_*$ (middle) and {\textbf{$M_*/R$}} as a $\sigma_*^2$ proxy (right panel). Radio-loud fraction exhibits and increasing trend with the observed axis ratio in all three panels.}
    \label{fig:heatmap}
\end{figure*}

\section{Which Parameters Correlate the Radio-Loud AGN Fraction?}
In the following subsections we explore the RL fraction trend with the $\sigma_*$ \citep[or equivalently BH mass;][]{vdbosch16} for all, quiescent and star-forming galaxies. We also explore the aspect of radio-loudness having a preference to occur more frequently among galaxies of a specific shape.

\subsection{Radio-Loud AGN Fraction Increases with $\sigma_*$}
 It has been shown in the local universe that the RL fraction of massive elliptical galaxies exhibits a strong increasing trend with $\sigma_*$ \citep{best05}. According to the BH mass$-\sigma_*$ relation \citep{vdbosch16} this implies a strong dependence of the RL fraction on the BH mass. In Figure~\ref{fig:rlfrac} (left panel) we show the fraction of quiescent galaxies in the selected sample (violet crosses), and the RL fraction among all (yellow circles), quiescent (red empty circles) and star-forming (blue empty circles) galaxies as a function of $\sigma_*$ and the corresponding BH mass \citep{vdbosch16}. It is evident that the RL fraction exhibits an increasing trend with $\sigma_*$, which reproduces the result from \cite{best05}. 
 
 Evidently, the RL fraction is increasing among star-forming galaxies as well, and these galaxies appear to host RL AGN as frequently as quiescent galaxies at fixed $\sigma_*$ \citep[i.e.][]{janssen12}. Since the star-forming galaxies host RL AGN at a comparable rate as the quiescent galaxies, then the radio-loudness is not a sufficient condition for a galaxy to become quiescent. %, however, it likely has an important role in the maintenance of quiescence.

\subsection{Geometry Plays a Role in Radio-Loud Fraction}
For the remainder of this letter we only consider quiescent galaxies, as we are interested in constraining maintenance-mode feedback.
We divide the sample into six evenly spaced bins of $\sigma_*$ and four in the projected axis ratio. In each bin we determine the fraction of RL galaxies. The right panel of Figure~\ref{fig:rlfrac} presents the heat-map of the projected axis ratio distribution as a function of $\sigma_*$ for quiescent galaxies, color-coded by the fraction of RL galaxies. %Two panels demonstrate RL fraction trends among the sample of all (left panel) and quiescent (right panel) and star-forming (right panel)
The top number inside of each bin corresponds to the number of RL galaxies, while the bottom number corresponds to the total number of galaxies. 
We observe that RL fraction among quiescent galaxies exhibits an increasing trend with both parameters -- $\sigma_*$ and the projected axis ratio. 

%The similar trend can be seen for all galaxies, however, it is likely driven by the trend seen for the quiescent population.
%\textcolor{red}{The increasing trend in the RL fraction with both parameters -- stellar velocity dispersion and the projected axis ratio, is best visible for quiescent galaxies (middle panel). The increasing RL fraction trend with the projected axis ratio seen for all galaxies (left panel) is likely driven by the trend seen for the quiescent population, since the RL fraction trend with the projected axis ratio is weak for star-forming galaxies.}

This result illustrates that the important parameters in determining the probability for a galaxy to ``switch on'' as a RL AGN is not only its $\sigma_*$ but also the geometry of the galaxy. 

We note that the $\sigma_*$ measurement for all but face-on galaxies is to a certain degree affected by contribution of the rotational velocity. Due to this effect, the actual $\sigma_*$ should be lower for inclined galaxies, further suggesting lower RL fractions. Assuming that the M$_*$ roughly corresponds to the virial mass, then the ratio between the galaxy M$_*$ and its inclination-independent semi-major axis (half-light) radius R, can serve as a proxy for the intrinsic $\sigma_*$ free of the rotational velocity effects, and independent on the inclination. % -- for quiescent galaxies we see an increasing RL fraction trend. , however, with star-forming galaxies the trend is affected by the low number statistics.
Figure~\ref{fig:heatmap} shows the RL fraction trend with the projected axis ratio for quiescent galaxies, in narrow bins of the $\sigma_*$ (left hand), M$_*$ (middle) and the $\sigma_*$ proxy log(M$_*$/R) (right panel). These trends are representative of the sample as a whole. %where R is the half light radius. 
%The increasing trend in all three panels implies that the RL fraction depends on the inclination of the galaxy, and likely on its intrinsic shape.
The increasing trend in all three panels is likely dominated by a dependence of the RL fraction on the intrinsic shape of the galaxy, rather then its inclination. We note that replacing the semi-major axis radius with the inclination-independent circularized radius in estimating the $\sigma_*$ proxy log(M$_*$/R) results in an even stronger trend than seen in the right panel of Figure~\ref{fig:heatmap}.
Therefore, correcting $\sigma_*$ measurement of inclined galaxies for the rotational velocity effects would not significantly alter the increasing trend we see, which implies the preference for RL AGN to occur in round galaxies.

%\begin{itemize}
%    \item add : the projected AR vs SFR shows no trend with the RL fraction
%\end{itemize}

%Since the geometry plays an important role in the RL fraction, then the increasing RL fraction trend among star-forming galaxies seen in Figure~\ref{fig:rlfrac} may be driven by the shape of the galaxy, meaning that the number of round star-forming galaxies at higher $\sigma_*$ values may dominate over the number of flat star-forming galaxies. 
%If so, this would imply the preference for RL AGN to occur in round galaxies. We test this hypothesis in the following subsection.

\begin{figure*}
    \centering
    \includegraphics[width=\linewidth]{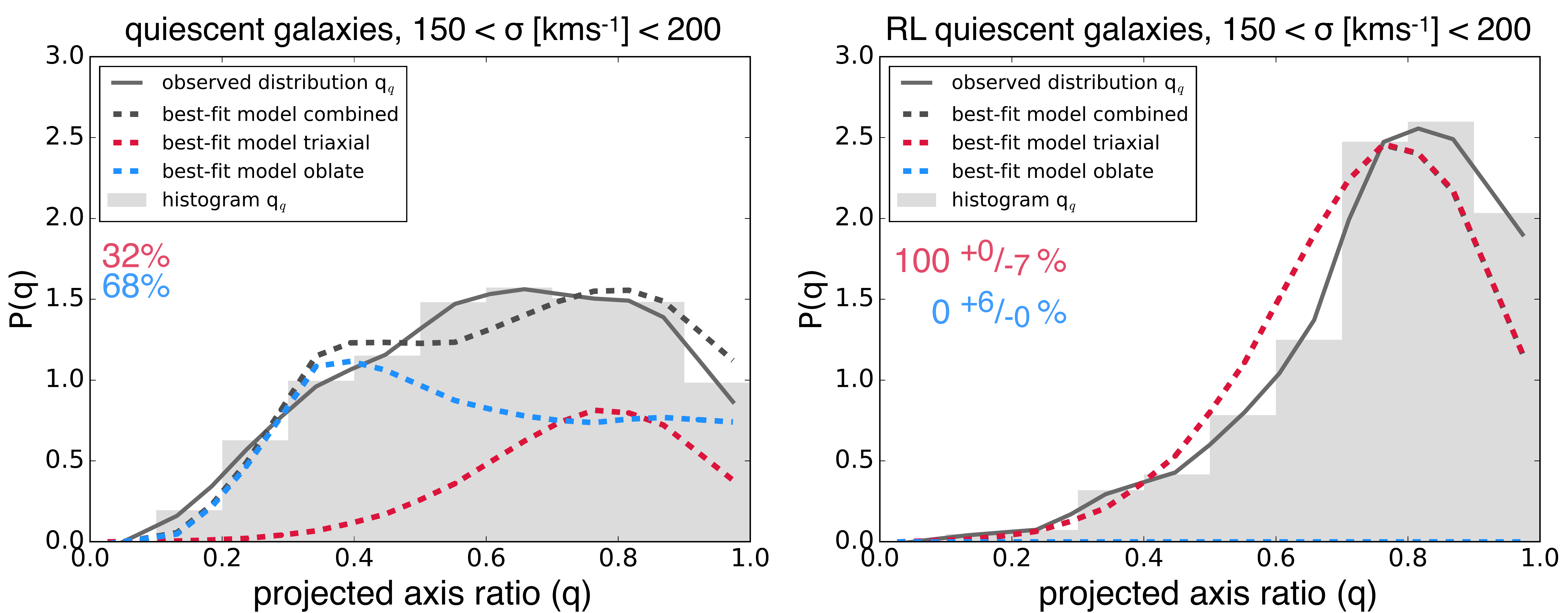}
    \caption{Normalized histogram of the observed axis ratio distribution (grey bins) for quiescent (left) and RL quiescent (right panel) galaxies and the function describing the observed axis ratio distribution (grey curve) in the $\sigma_*$ bin 150 $-$ 200 kms$^{-1}$. We also show the best-fit two-component model probability distribution (black dashed curve), and the underlying triaxial (red dashed curve) and oblate (blue dashed curve) best-fit fractions.}
    \label{fig:histogram}
\end{figure*}

\subsection{Radio-Loud AGN Fraction of Intrinsically Round and Flat Galaxies}
\cite{chang13} developed a two-population model to reproduce the observed axis ratio distribution of quiescent galaxies up to $z=2.5$, combining round (triaxial) and flat (oblate) populations of galaxies. They perform their modelling in three different mass bins, and to simplify the analysis we only use the fit parameters of the highest stellar mass bin (M$_*$ $>$ 10.8)
\footnote{This mass limit corresponds to a $\sigma_*$ limit of $\sim 120\rm{km~s}^{-1}$, but even below this limit the shape parameters are not very different and our analysis would not be affected.}  where the intrinsic axis ratio of the oblate component is b\footnote{b = intermediate axis of the triaxial ellipsoid} = 0.29$\pm$0.07. %to reconstruct the model two-population intrinsic axis ratio distribution of local galaxies. 
The best-fit triaxility (T\footnote{T = [1 - $\beta ^{2}$]/[1 - $\gamma ^{2}$]} = 0.64$\pm$0.06) and ellipticity (E\footnote{E = 1 - $\gamma$} = 0.41$\pm$0.02) parameter values, and their standard deviations ($\sigma_T$, $\sigma_E$ = 0.08$\pm$0.05, 0.19$\pm$0.02) are used to generate distributions of intrinsic intermediate-to-long axis ratio $\beta$ and short-to-long axis ratio $\gamma$. 
Combining these distributions with random viewing angles provide us with model triaxial and oblate intrinsic axis ratio distributions. 
%Generated $\beta$ and $\gamma$ distributions provide us with model triaxial and oblate intrinsic axis ratio distributions. 
%For simplicity in analysis in the following subsection we assume that the described two population model which was built for a sample of quiescent galaxies, can be applied to the total sample and to the sample of star-forming galaxies.

%\subsection{RL galaxies prefer round galaxies}
%\begin{itemize}
%    \item add how you turned the histogram into the function and that you compared the obtained function to the lin. comb. of trix and obl to get the best fit (gray curve)
%\end{itemize}
We split the sample of quiescent galaxies into six equally spaced $\sigma_*$ bins, and in each bin we find the best-fit linear combination of model triaxial and oblate shape distributions which reproduce the observed axis ratio distribution. This gives us the triaxial and oblate fraction of galaxies in the specific $\sigma_*$ bin. As an example, in Figure~\ref{fig:histogram} we show a normalized histogram (gray bins) of the observed axis ratio (q) distribution, and a corresponding function describing the distribution obtained after binning, for quiescent (left panel) and for RL quiescent (right panel) galaxies in the range of $\sigma_*$ values from 150 - 200 km s$^{-1}$. For illustration, we choose the $\sigma_*$ bin in which the total population is similarly populated by both triaxial and oblate galaxies according to the two-population model. Blue and red dashed curves represent best-fit oblate and triaxial fractions, respectively. We see that the triaxial fraction (100 $\pm$ $^{0}_{7}$\%) of RL AGN in the observed quiescent population in this $\sigma_*$ bin completely dominates over the oblate fraction, implying a more frequent occurrence of RL AGN among round galaxies. 

The best-fit model triaxial and oblate fraction for the observed axis ratio distribution of quiescent galaxies can be rescaled to fit the observed axis ratio distribution of RL quiescent galaxies. Multiplying these rescaling factors by the RL fraction gives the RL fraction among triaxial and oblate galaxies. 

\begin{figure}
    \centering
    \includegraphics[width=\linewidth]{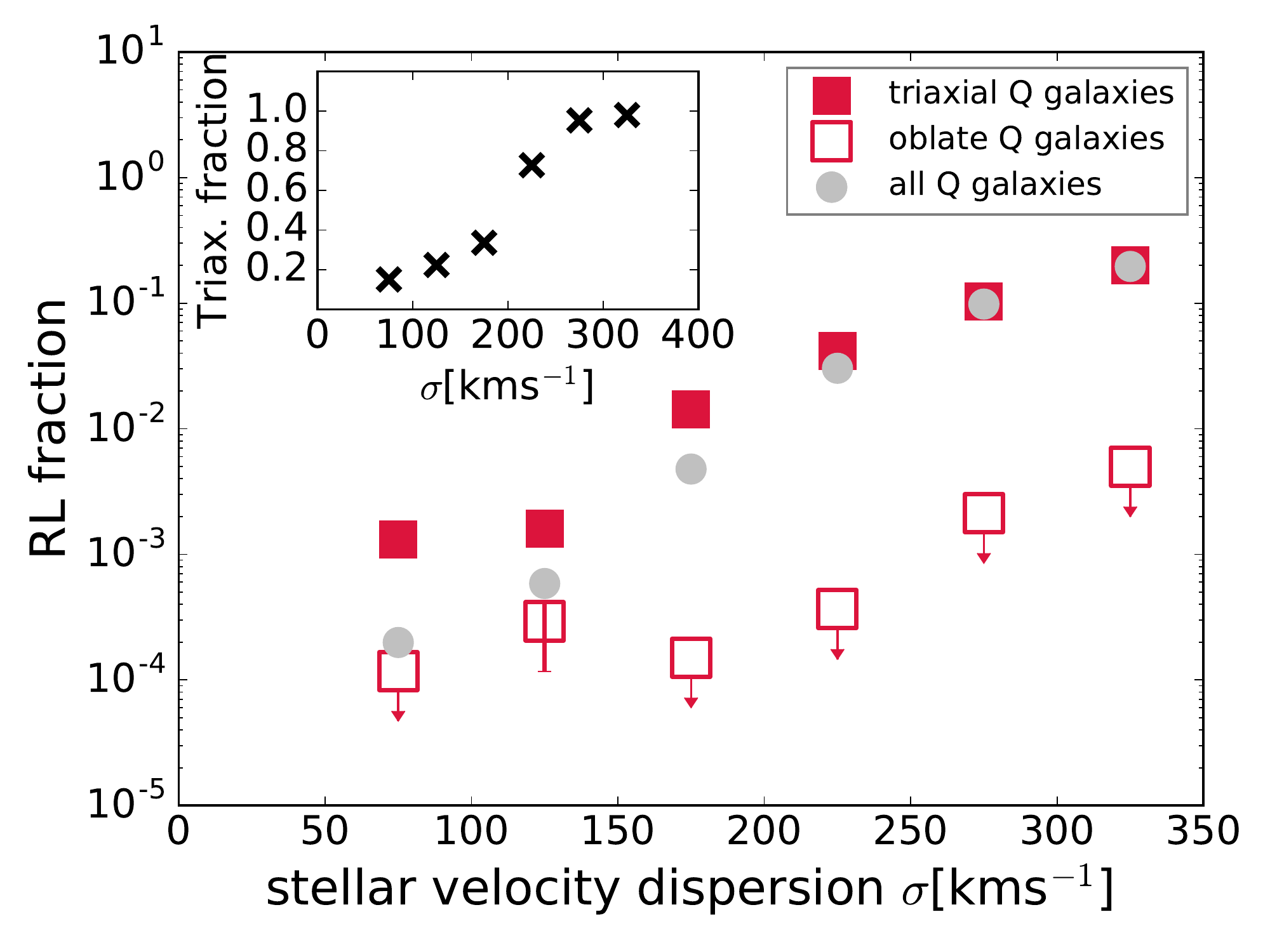}
    \caption{The RL fraction among triaxial (red squares) and oblate (open red squares) galaxies as a function of the $\sigma_*$ for quiescent galaxies. The RL fraction is higher among triaxial galaxies. The plot also shows the RL fraction among quiescent galaxies (grey circles), and the best-fit triaxial fraction (black crosses).}
    \label{fig:trix}
\end{figure}

Figure~\ref{fig:trix} shows the RL fraction among triaxial (red filled squares) and oblate (red open squares) galaxies for quiescent population as a function of the $\sigma_*$. The overall triaxial fraction (black crosses) increases with the $\sigma_*$. The increase in RL fraction with $\sigma_*$ is entirely due to its increase among the triaxial population. In fact, the inferred RL fraction among oblate population is essentially zero. This analysis, by accounting for round (face-on) disk-like galaxies, further strengthens our result that RL AGN preferentially reside in intrinsically round galaxies. We note that we do not claim that the RL fraction among disk-like quiescent galaxies is, in fact, zero; the triaxial population is defined to consist of objects with a wide range in intrinsic $\beta$ and $\gamma$ values (see above), including a subset with $\beta$ $\sim$1 and small $\gamma$, that is, disk-like geometries. 
%Following \citep{fanidakis11} and the spin paradigm: since the SMBH in round elliptical galaxies are most likely rapid rotators -- presumably formed via gas-poor major-merger events, the implication of this result is that the BH spin is an important factor in determining the occurrence and strength of RL AGN in galaxies.

\subsection{Satellites, Centrals and Halo Mass}
We now investigate whether the correlation between the shape and RL fraction is driven by an underlying connection between dark matter halo properties and RL AGN. This may be expected in case disk-like galaxies live in less massive halos than round galaxies, or if satellite galaxies, which are more often disk-like than centrals \citep{vdwel10}, are less likely to be RL AGN. 

%For this purpose we cross-matched our sample with the the SDSS DR7 based group catalog created by \citep{yang07}, using their "petroC"\footnote{http://gax.sjtu.edu.cn/data/Group.html}. We focus on quiescent galaxies with $\sigma_*$ in the range $200<\sigma_* / (\rm{km}~\rm{s^{-1}}) < 230$, which shows a strong correlation between the shape and RL AGN incidence, and for which the satellite fraction is non-neglible \citep[$\sim20$\%,][]{vdbosch08}. We find that the RL AGN fractions for satellites and centrals are not different: 2.3\% for the satellites, and 2.8\% for the centrals. Moreover, the trends with galaxy shape for centrals and satellites are consistent with each other. Finally, there is no substantial difference ($<$0.1 dex) between the halo masses of round central galaxies and flat central galaxies.
We cross-match our sample with the  SDSS DR7 based "petroC"\footnote{ http://gax.sjtu.edu.cn/data/Group.html } group catalog created by \cite{yang07}. For this halo catalog, centrals are chosen based on stellar mass, and halo masses are assigned to groups based on their total stellar mass. We focus on quiescent galaxies with  $200<\sigma_* / (\rm{km}~\rm{s^{-1}}) < 230$, which shows a strong correlation between the shape and RL AGN incidence, and for which the satellite fraction is non-neglible \citep[$\sim20$\%,][]{vdbosch08}. We find that RL fractions are similar: 2.3\% for the satellites, and 2.8\% for the centrals, and their trends with galaxy shape are consistent with each other. There is no substantial difference ($<$0.15 dex) between the halo masses of round central galaxies and flat central galaxies, and selecting galaxies by halo mass still contains a strong correlation between RL AGN fraction and galaxy shape.

We conclude that the strong dependence of RL AGN incidence on galaxy shape is intrinsic and cannot be explained by differences between RL AGN fractions in satellite and central galaxies, nor to differences in halo mass.

\section{Discussion}

In this paper we demonstrate that RL AGN occur more frequently in round than in flat quiescent galaxies at any M$_*$ or $\sigma_*$. The analysis of the projected axis ratio distribution of RL AGN hosted by quiescent galaxies shows that the increased RL fraction at higher $\sigma_*$ values is entirely driven by intrinsically round (triaxial) galaxies. The RL fraction reaches $\sim$20\% for the most massive, round quiescent galaxies ($\sigma_*$ $\sim$300 km s$^{-1}$), whereas geometrically flat (disk-like) quiescent galaxies have an extremely low RL AGN fraction ($<$ $\sim$10$^{-3}$), even at high $\sigma_*$ ($>$200 km s$^{-1}$).

This result is consistent with the common notion that luminous radio AGN reside in massive, elliptical galaxies. However, thus far, the correlation between the RL AGN occurrence in quiescent galaxies and a second global property, besides its $\sigma_*$ (BH mass), has not been explicitly demonstrated. The roundness of a galaxy likely reflects its merger history, and the connection between an active merger history and the occurrence of low-accretion rate radio AGN has been made by many authors \citep{ruiter05, capetti06, sikora07, fanidakis11}. The physical basis for this connection has been referred to as the ``spin paradigm'' \citep{blandford77, wilson95, hughes03, mckinney04} in which only BHs with large spin parameters can launch radio jets. In turn, a large spin parameter is the result of BH-BH mergers that reflect the merger history of the host galaxy.

Regardless of the physical explanation, the lack of RL AGN in disk-like quiescent galaxies raises important questions. In galaxy formation and evolution models, low-accretion rate AGN are universally assumed to play a crucial role in preventing excessive SF in massive galaxies, independent of their structure. Our findings suggest that, if this model is generally valid, the AGN feedback does not manifest itself at radio wavelengths for galaxies with disk-like structures. We conclude that there is compelling evidence for maintenance-mode feedback for intrinsically round galaxies, but that further study is needed to bolster the evidence for such feedback as a general explanation for the lack of star-formation activity in massive galaxies.

\section{Acknowledgements}
We thank the anonymous referee for providing useful feedback. 
Funding for the SDSS and SDSS-II has been provided by the Alfred P. Sloan Foundation, the Participating Institutions, the National Science Foundation, the U.S. Department of Energy, the National Aeronautics and Space Administration, the Japanese Monbukagakusho, the Max Planck Society, and the Higher Education Funding Council for England\footnote{http://www.sdss.org/}.
This project has received funding from the European Research Council (ERC) under the European Union's Horizon 2020 research and innovation programme (grant agreement 683184).
GvdV acknowledges funding from the European Research Council (ERC) under the European Union's Horizon 2020 research and innovation programme under grant agreement No 724857 (Consolidator Grant ArcheoDyn)
  
%\bibliographystyle{aasjournal}
%\bibliography{sdss.bib}

\end{document}